%% file: main.tex
\documentclass[conference]{IEEEtran}
\IEEEoverridecommandlockouts 
\usepackage[utf8]{inputenc}
\usepackage[T1]{fontenc}
\usepackage{cite}
\usepackage{graphicx}
\usepackage{float}
\usepackage{booktabs}
\usepackage{tabularx}
\usepackage{balance}
\usepackage{amsmath,amssymb,amsfonts,mathtools}
\usepackage{tikz}
\usetikzlibrary{math}
\usepackage{algorithm}
\usepackage{algpseudocode}
\usepackage[caption=false,font=footnotesize,labelfont=sf,textfont=sf]{subfig}
\usepackage{xcolor}
\usepackage[nolist,nohyperlinks]{acronym}
\usepackage{setspace}
\usepackage{stfloats}

\let\Algorithm\algorithm
\renewcommand\algorithm[1][]{\Algorithm[#1]\setstretch{1.1}}

\input{acronyms.tex}

\begin{document}

\title{OFDM-Based ISAC Imaging of Extended Targets via Inverse Virtual Aperture Processing}
\author{Michael~Negosanti, Lorenzo~Pucci, and Andrea~Giorgetti
\thanks{This work was supported, in part, by the CNIT National Laboratory WiLab and the WiLab-Huawei Joint Innovation Center, and in part by the European Union under the Italian National Recovery and Resilience Plan (NRRP) of NextGenerationEU, partnership on ``Telecommunications of the Future'' (PE00000001 - program ``RESTART'').}
\thanks{The authors are with the Dept. of Electrical, Electronic, and Inf. Eng. ``Guglielmo Marconi,'' University of Bologna, and CNIT/WiLab, Italy (e-mail: \{michael.negosanti2, lorenzo.pucci3, andrea.giorgetti\}@unibo.it).}
}

\maketitle

\begin{abstract}
This work investigates the performance of an \ac{ISAC} system exploiting \ac{IVA} for imaging moving extended targets in vehicular scenarios. A \ac{BS} operates as a monostatic sensor using \acs{MIMO}–\acs{OFDM} waveforms. Echoes reflected by the target are processed through motion-compensation techniques to form an \ac{IVA} range–Doppler (cross-range) image. A case study considers a 5G NR waveform in the upper mid-band, with the target model defined in 3GPP Release 19, representing a vehicle as a set of spatially distributed scatterers. Performance is evaluated in terms of \ac{IC} and the \ac{RMSE} of the estimated target-centroid range. Finally, the trade-off between sensing accuracy and communication efficiency is examined by varying the subcarrier allocation for \ac{IVA} imaging. The results provide insights for designing effective sensing strategies in next-generation radio networks.
\end{abstract}
\acresetall


\section{Introduction}
\IEEEPARstart{T}{he} \ac{ISAC} paradigm is widely recognized as one of the key pillars of next-generation 6G networks, enabling future systems to move beyond traditional communication services and provide ubiquitous sensing capabilities. Within the 6G vision, such native sensing is expected to support a wide range of applications, including assisted navigation, activity recognition, motion tracking, environmental monitoring, and imaging of objects and their surroundings. Among the sensing methods best suited to these applications, particularly for autonomous driving, which requires operation under all weather conditions and high-resolution mapping, virtual-aperture-based techniques are especially promising \cite{Tagliaferri1,Tagliaferri2,Tagliaferri3}.

In this context, the sensing capabilities of \ac{OFDM} waveforms can also be leveraged for radio imaging applications based on \ac{VA} or \ac{IVA} techniques \cite{Bi-ISAR, JCASAR, ISAC-ISAR-USRP}. In particular, \ac{IVA} stands out as one of the most suitable technologies, as it enables the imaging of non-cooperative targets even when the sensing \ac{BS} remains stationary \cite{chen2014inverse}. To the best of the authors’ knowledge, the existing literature still lacks studies analyzing an \ac{OFDM}-based \ac{ISAC}–\ac{IVA} system in a 3D scenario, aimed at assessing its performance and the impact of resource allocation between imaging and communication.


This study investigates the use of \ac{IVA} imaging in \ac{MIMO}–\ac{OFDM} systems, such as 5G and 6G, where the sensing \ac{BS} operates in a monostatic configuration. The objective is to assess the performance of \ac{IVA}-based methods in terms of image contrast and target-centroid estimation accuracy as a function of the sensing resources allocated. In particular:
$i)$ We design an \ac{IVA}-based imaging system at the \ac{BS}, leveraging \ac{MIMO}–\ac{OFDM} waveforms in a monostatic configuration. $ii)$ We adopt the extended-target model recently standardized in 3GPP Release 19 to represent a vehicle. $iii)$ We apply motion-compensation techniques to generate an \ac{IVA} range–Doppler image, which is then post-processed through thresholding to estimate the target-centroid range. $iv)$ We analyze how the fraction of subcarriers allocated to sensing affects both the \ac{RMSE} of the target-centroid range estimate and the \ac{IC} metric, considering a case study based on 5G NR operation in the upper mid-band.

Throughout the paper, capital boldface letters denote matrices; lowercase boldface letters denote vectors; $(\cdot)^\mathsf{T}$, $(\cdot)^\mathsf{H}$, $\lceil\cdot\rceil$, $\|\cdot\|$, and $\|\cdot\|_F$ represent transpose, conjugate transpose, ceiling function, Euclidean, and Frobenius norms, respectively; $\mathbf{X}_{:,n}$ is the $n$th column of $\mathbf{X}$; $\mathbf{1}_{M \times N}$ and $\mathbf{1}_N$ are all-one matrix and vector, respectively; $\mathbb{E}\{\cdot\}$ denotes expectation; $\mathbf{x}\sim\mathcal{CN}(\mathbf{0},\boldsymbol{\Sigma})$ indicates a zero-mean circularly symmetric complex Gaussian vector with covariance $\boldsymbol{\Sigma}$; $\operatorname{FFT}_N\{\cdot\}$ denotes an $N$-point \ac{FFT}.


\section{System Model}\label{sec:intro}
\begin{figure}[t]
    \centering
    \includegraphics[width=1.4\columnwidth]{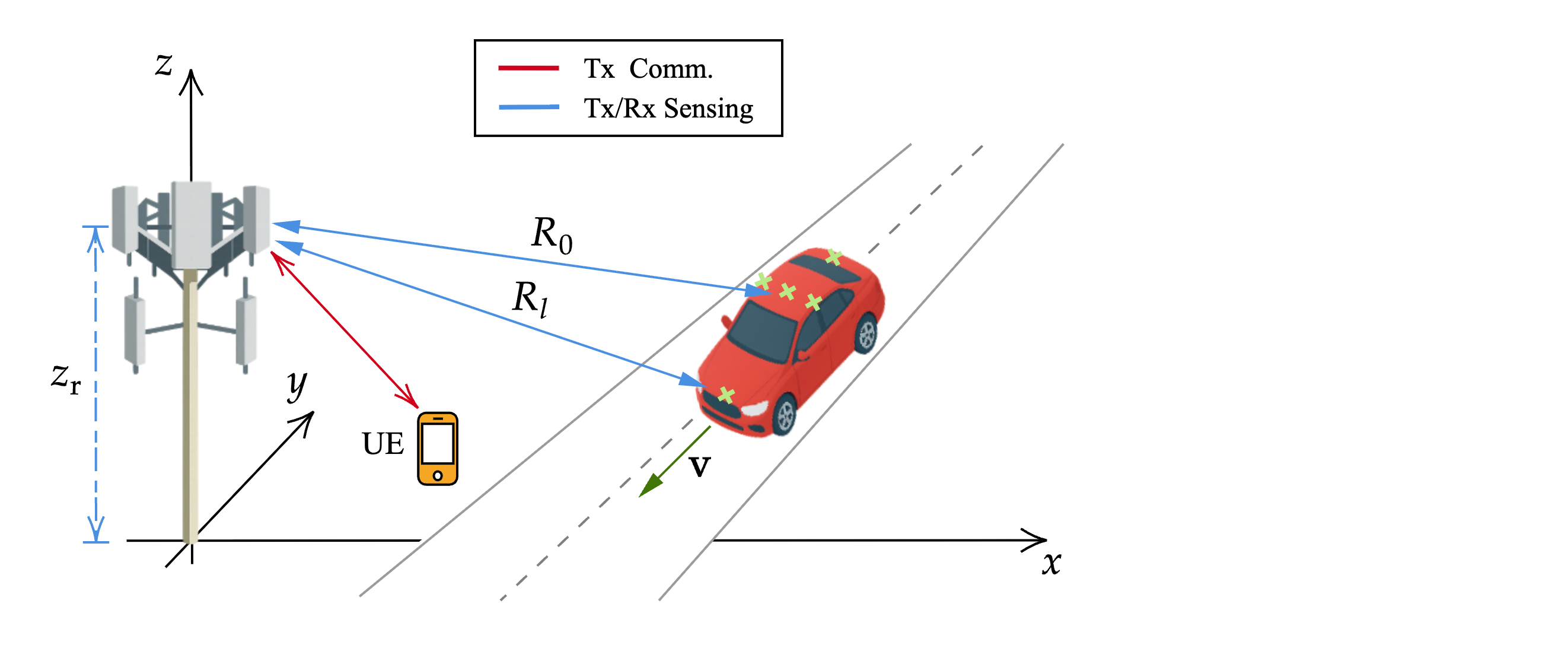}
    \caption{Schematic representation of an ISAC–\ac{IVA} setup with a \ac{BS} using part of the physical-layer resources to perform imaging of a moving vehicle.}
    \label{fig:schematic}
\end{figure}

We consider a monostatic \ac{MIMO}-\ac{OFDM} \ac{BS} as shown in Fig.~\ref{fig:schematic}, that communicates with \acp{UE}, while simultaneously processing echoes generated by a moving vehicle to produce \ac{IVA} images.
The \ac{BS} is equipped with a transmit \ac{ULA} with $N_\mathrm{T}$ antennas for downlink transmission and a separate receive \ac{ULA} with $N_\mathrm{R}$ elements for backscattered signal collection, both horizontally aligned along the $y$-axis of Fig.~\ref{fig:schematic}. For both the \acp{ULA}, we consider an inter-element spacing $d=\lambda/2$, where $\lambda=c/f_\mathrm{c}$, with $c$ the speed of light and $f_\mathrm{c}$ the carrier frequency.

\subsection{Transmit signal}\label{sec:TX_sig}

%
The \ac{ISAC} system transmits \ac{OFDM} frames composed of $M$ \ac{OFDM} symbols in the time domain and $K$ active subcarriers. Given a subcarrier spacing $\Delta f = 1/T$, where $T$ denotes the \ac{OFDM} symbol duration (without the cyclic prefix), the total signal bandwidth is $B = K \Delta f$. The total frame duration is $T_\mathrm{f} = M T_\mathrm{s}$, where $T_\mathrm{s} = T + T_\mathrm{g}$ is the total \ac{OFDM} symbol duration including the \ac{CP} of duration $T_\mathrm{g}$, introduced to avoid \ac{ISI}.
For each frame, a portion of the time–frequency resources is allocated to sensing (imaging) operations, with a \ac{SRI}, i.e., the time between two consecutive sensing symbols, equal to $T_\mathrm{SRI} = N_\mathrm{s} T_\mathrm{s}$, where $N_\mathrm{s} \in \mathbb{N}$, and a sensing bandwidth $B_\mathrm{s} = K_\mathrm{s} \Delta f$. To illuminate the entire spatial extent of a moving target, $M_\mathrm{s}$ sensing \ac{OFDM} symbols are considered over multiple consecutive frames, leading to a total \ac{CPI} of $T_\mathrm{CPI} = M_\mathrm{s} T_\mathrm{SRI} > T_\mathrm{f}$.

Hereinafter, we indicate with $\rho_\mathrm{f}=K_\mathrm{s}/K$ and $\rho_\mathrm{t} = \tfrac{T_\mathrm{s}\lceil M/N_\mathrm{s}\rceil}{T_\mathrm{f}}$ the fraction of subcarriers and fraction of frame-time allocated to sensing, respectively.

For each \ac{OFDM} frame, a time-frequency grid $\mathbf{X}$ of complex modulation symbols $x_{k,m}$ is considered, where each symbol corresponds to subcarrier index $k=0,\dots,K-1$ and time index $m=0,\dots,M-1$, and is normalized such that $\mathbb{E}\{|x_{k,m}|^2\}=1$. Prior to transmission, the symbols are precoded across the $N_\mathrm{T}$ transmit antennas by means of a digital beamforming vector $\mathbf{w}_\mathrm{T} \in \mathbb{C}^{N_\mathrm{T} \times 1}$, which depends on the system operation mode, i.e., sensing or communication. This work focuses on the sensing operation, where the goal is to perform moving-target imaging via \ac{IVA} processing. Therefore, details related to communication beamforming are omitted.

For imaging applications, $\mathbf{w}_\mathrm{T}$ is designed to generate a wide sensing beam, with beamwidth $\Delta\Theta_\mathrm{T}$, to cover the target trajectory during a \ac{CPI}, following the approach in \cite{FriedlanderMIMOradar}. Although this design entails a reduction in beamforming gain compared to a narrow beam, it ensures that the entire spatial extent of the target is illuminated with approximately constant gain and phase.\footnote{Each wide beam is, in principle, designed to illuminate a single target along its trajectory. In the presence of multiple targets, they can be sequentially illuminated, e.g., in a time-division manner.} 
Preserving phase coherence during target motion is essential for accurate inverse aperture synthesis and high-fidelity image reconstruction. The beamforming vector is normalized as $\|\mathbf{w}_\mathrm{T}\|^2 = P_\mathrm{avg}$, where $P_\mathrm{avg} = P_\mathrm{T}/K$ denotes the average transmit power per subcarrier, given a total transmit power $P_\mathrm{T}$.
After beamforming, the $[N_\mathrm{T}\times 1]$ vector of complex transmit symbols $\mathbf{x}[k,m] = x_{k,m} \mathbf{w}_\mathrm{T}$ is obtained for each subcarrier $k$ and \ac{OFDM} symbol $m$.
Hereinafter, the index $m_\mathrm{s} \in \{0, \dots, M_\mathrm{s}-1\}$ is introduced to indicate sensing \ac{OFDM} symbols extracted from the transmitted grids over multiple consecutive frames with a sampling interval $T_\mathrm{SRI}$. 



\subsection{Sensing signal received}
At the sensing receiver, echoes reflected by surrounding targets are collected. After demodulation through the \ac{FFT}, the sensing receiver extracts the symbols over the $K_\mathrm{s}$ contiguous subcarriers every $T_\mathrm{SRI}$ for further processing. For subcarrier $k$ and sensing time index $m_\mathrm{s}$, the complex received signal vector of size $[N_\mathrm{R} \times 1]$, under the assumption of negligible \ac{ISI} and \ac{ICI}, is expressed as\footnote{Note that monostatic sensing requires a full-duplex architecture with \ac{SI} cancellation. In this work, we consider the residual \ac{SI} after cancellation to be negligible with respect to Gaussian noise.}
\begin{equation}
    \mathbf{y}[k,m_\mathrm{s}] = \mathbf{H}[k,m_\mathrm{s}] \mathbf{x}[k,m_\mathrm{s}] + \mathbf{n}[k,m_\mathrm{s}]
    \label{eq:y_tilde}
\end{equation}
where $\mathbf{H}[k,m_\mathrm{s}] \in \mathbb{C}^{N_\mathrm{R} \times N_\mathrm{T}}$ denotes the \ac{MIMO} channel matrix containing information related to the targets, and $\mathbf{n}[k,m_\mathrm{s}] \sim \mathcal{CN}(0,\sigma^2\mathbf{I}_{N_\mathrm{R}})$ represents the noise, assumed \ac{i.i.d.} across subcarriers and time. The noise variance is $\sigma^2 = N_0 \Delta f$, with $N_0$ the one-sided noise \ac{PSD}.

We consider \ac{LoS} propagation and a single extended target composed of $L$ scattering points. For a given scatterer $l$, the round-trip delay, azimuthal \ac{AoA}/\ac{AoD}, and complex channel gain sampled at time $m_\mathrm{s}T_\mathrm{SRI}$ are denoted by $\tau_l[m_\mathrm{s}]\triangleq\tau_l(m_\mathrm{s}T_\mathrm{SRI})$, $\theta_l[m_\mathrm{s}]\triangleq\theta_l(m_\mathrm{s}T_\mathrm{SRI})$, and $\alpha_l[m_\mathrm{s}]\triangleq\alpha_l(m_\mathrm{s}T_\mathrm{SRI})$, respectively. These parameters are assumed constant within each \ac{OFDM} symbol (fast time) but may vary across sensing symbols (slow time) due to target motion. Under these assumptions, the channel matrix in \eqref{eq:y_tilde} can be written as
\begin{equation}\label{eq:channel-matrix}
    \mathbf{H}[k,m_\mathrm{s}] 
= \sum_{l = 1}^{L} 
    \underbrace{\alpha_l[m_\mathrm{s}] e^{-\imath 4\pi k \Delta f\, \frac{R_l[m_\mathrm{s}]}{c}}}_{\triangleq \beta_l[m_\mathrm{s}]} 
    \mathbf{a}_\mathrm{R}(\theta_l[m_\mathrm{s}])\mathbf{a}^\mathsf{H}_\mathrm{T}(\theta_l[m_\mathrm{s}])
\end{equation}
where $R_l[m_\mathrm{s}]$ denotes the distance between scatterer $l$ and the \ac{BS} at time $m_\mathrm{s}$ while $\mathbf{a}_\mathrm{R}(\cdot)$ and $\mathbf{a}_\mathrm{T}(\cdot)$ represent the receive and transmit array response vectors, respectively. The transmit array response is defined as \cite{van2002optimum}\footnote{%
Since the considered \acp{ULA} scan only the azimuth plane and the target elevation angle varies slightly during its motion, elevation effects are negligible and thus omitted from the array model.}
\begin{equation}
\mathbf{a}_\mathrm{T}(\theta_l[m_\mathrm{s}]) 
= \big[e^{-\imath\frac{N_\mathrm{T}-1}{2}\pi \sin(\theta_l[m_\mathrm{s}])}, \dots, e^{\imath\frac{N_\mathrm{T}-1}{2}\pi \sin(\theta_l[m_\mathrm{s}])}\big]^\mathsf{T}.
\end{equation}
The same definition applies to $\mathbf{a}_\mathrm{R}(\theta_l[m_\mathrm{s}])$ with $N_\mathrm{R}$  elements.

The complex channel gain models attenuation and phase shift along the $l$th path at time $m_\mathrm{s}$ and is given by $\alpha_l[m_\mathrm{s}] = \tilde{\alpha}_l[m_\mathrm{s}] e^{\imath \gamma_l[m_\mathrm{s}]}$. According to the radar equation, we define the channel gain as $\tilde{\alpha}_l[m_\mathrm{s}]= \sqrt{\frac{G_\mathrm{T}G_\mathrm{R}\sigma_l \, c^2}{(4 \pi)^3 \, f_\mathrm{c}^2 \, R_{l}^4[m_\mathrm{s}]}}$, where $G_\mathrm{T}$ and  $G_\mathrm{R}$ represent the single transmit and receive antenna gains, and $\sigma_l$ is the \ac{RCS} of the $l$th scatterer \cite{richards}. The phase is given by $\gamma_l[m_\mathrm{s}] = -4\pi f_\mathrm{c}\frac{R_l[m_\mathrm{s}]}{c} + \phi_l$, where the former term is a deterministic component due to the round-trip propagation delay, while $\phi_l\sim \mathcal{U}(0,2\pi)$ is a random scattering phase, representing the residual phase offset associated with the small-scale structure of the scatterer, assumed constant over the \ac{CPI}. 

Beamforming is then applied to the received symbol vector in \eqref{eq:y_tilde} by using the receive weight vector $\mathbf{w}_\mathrm{R} \in \mathbb{C}^{N_\mathrm{R}\times 1}$ to obtain the received symbol $y_{k,m_\mathrm{s}}$. The beamforming vector $\mathbf{w}_\mathrm{R}$ is assumed to be designed the same way as $\mathbf{w}_\mathrm{T}$, i.e., to produce a wide beam, with beamwidth $\Delta\Theta_\mathrm{R}$, centered at a given sensing azimuthal direction $\theta_\mathrm{s}$, with $\|\mathbf{w}_\mathrm{R}\|^2=1$ (see Section~\ref{sec:TX_sig}). 
Considering \eqref{eq:y_tilde}, and plugging into it the channel matrix in \eqref{eq:channel-matrix}, the resulting received symbol becomes
\begin{align}
\label{eq:y_beamformed}
y_{k,m_\mathrm{s}}
&= \left( \sum_{l=1}^L \beta_l[m_\mathrm{s}] \Upsilon(\theta_l[m_\mathrm{s}], \theta_\mathrm{s}) \right) x_{k,m_\mathrm{s}} + \tilde{n}_{k,m_\mathrm{s}}
\end{align}
where $\tilde{n}_{k,m_\mathrm{s}} = \mathbf{w}_R^{\mathsf{H}}\mathbf{n}_{k,m_\mathrm{s}} \sim \mathcal{CN}(0, \sigma^2)$ is the noise after beamforming, and $\Upsilon(\theta_l[m_\mathrm{s}], \theta_\mathrm{s}) \in \mathbb{C}$ is the composite transmitting-receiving beamforming gain along the azimuthal direction $\theta_l[m_\mathrm{s}]$ at time $m_\mathrm{s}$. Moreover, $\beta_l[m_\mathrm{s}]$, defined in \eqref{eq:channel-matrix}, contains information about the range and Doppler of scatterer $l$ at time $m_\mathrm{s}$. From \eqref{eq:y_beamformed}, we remove the transmitted data symbol $x_{k,m_\mathrm{s}}$, which represents a nuisance parameter, by element-wise reciprocal filtering \cite{Braun}. This operation is repeated for each $(k, m_\mathrm{s})$ pair to produce the sensing data matrix $\mathbf{G}_\mathrm{s} \in \mathbb{C}^{K_\mathrm{s} \times M_\mathrm{s}}$, which is subsequently used to reconstruct an \ac{IVA} range–Doppler (or range--cross-range) image. In particular, the $(k,m_\mathrm{s})$ entry of $\mathbf{G}_\mathrm{s}$ is
\begin{equation}
    g_{k,m_\mathrm{s}} = \frac{y_{k,m_\mathrm{s}}}{x_{k,m_\mathrm{s}}}\\
     =  \sum_{l=1}^L \beta_l[m_\mathrm{s}] \Upsilon(\theta_l[m_\mathrm{s}], \theta_\mathrm{s}) + \nu_{k,m_\mathrm{s}}
    \label{eq:G_matrix}
\end{equation}
where $\nu_{k,m_\mathrm{s}} = \tilde{n}_{k,m_\mathrm{s}}/x_{k,m_\mathrm{s}}\sim\mathcal{CN}\Bigr(0,\sigma^2 \mathbb{E}\left\{1/|x_{k,m_\mathrm{s}}|^2\right\}\Bigl)$.\footnote{When a constant-envelope modulation is used (e.g., \ac{QPSK}), $|x_{k,m_\mathrm{s}}|^2 = 1$ for all $k,m_\mathrm{s}$, hence $\mathbb{E}\left\{1/|x_{k,m_\mathrm{s}}|^2\right\} = 1$. Otherwise, the noise variance increases after element-wise division (see \cite{PucArcGio:PP25}).}

\subsection{Range and Doppler reconstruction}
As shown in Fig.~\ref{fig:schematic}, we consider a vehicular target moving with uniform rectilinear motion at a constant speed $v = |\mathbf{v}|$, where $\mathbf{v}$ denotes the velocity vector. The target’s motion alters its aspect angle as observed by the sensing system. This variation in viewing angle results in an apparent rotation of the target, which is effectively equivalent to a true rotational motion from an imaging perspective. The resulting apparent rotation induces spatially dependent Doppler frequency shifts that can be exploited to discriminate scatterers located at different spatial positions \cite{chen2014inverse}.

Let $R_0$ denote the initial reference range between the center of the extended target and the \ac{BS}, and let $(x,y,z)$ and $(x',y',z')$ represent the global (space-fixed) and local (target-fixed) Cartesian coordinate systems, respectively. The space-fixed frame is centered at the \ac{BS} position with zero rotation angle, whereas the target-fixed frame has its origin at the target center. Assuming that the \ac{BS}-to-target distance is much larger than the physical dimensions of the target, which is commonly referred to as the straight iso-range approximation \cite{RadarMaritime}, the 3D Euclidean distance $R_l$ between the \ac{BS} and the $l$th scatterer at slow-time index $m_\mathrm{s}$ is given by
\begin{align}\label{eq:R}
R_l[m_\mathrm{s}] \approx {}& R_\mathrm{c}[m_\mathrm{s}] + \bigl[x_l' \cos\!\bigl(\Theta_\mathrm{c}[m_\mathrm{s}]-\theta_0\bigr) \\
& {}- y_l' \sin\!\bigl(\Theta_\mathrm{c}[m_\mathrm{s}]-\theta_0\bigr)\bigr]\cos\varphi+ z_l' \sin\varphi\nonumber
\end{align}
%
%
%
%
%
where $(x'_l, y'_l, z'_l)$ are the local coordinates of scatterer $l$, $\theta_0$ is the initial target-centroid azimuth angle, and $\varphi$ is the elevation of the \ac{LoS} path between the \ac{BS} and the target center. The translational displacement of the target geometric center is modeled as $R_\mathrm{c}[m_\mathrm{s}]=R_0+v_0 {m_\mathrm{s}}T_{\mathrm{SRI}} + \frac{1}{2}a_0 ({m_\mathrm{s}}T_{\mathrm{SRI}})^2$, where $v_0$ is the initial radial velocity, and $a_0 \;=\; \big(\omega_0 \cos \varphi\big)^2 R_0$ is the centripetal acceleration induced by the apparent rotation of the target, with $\omega_0\cos\varphi$ representing the effective angular velocity in the \ac{IPP}.
Moreover, $\Theta_\mathrm{c}[m_\mathrm{s}] = \Theta_0 + \omega_0 {m_\mathrm{s}}T_{\mathrm{SRI}}$ is the rotation angle,
with $\Theta_0$ the initial rotation angle (here equivalent to the target heading direction angle). The effective angular velocity $\omega_0\cos\varphi$ is assumed constant over the \ac{CPI}.

Applying the angle-difference identities and a Taylor expansion of $\sin(\omega_0 m_\mathrm{s} T_{\mathrm{SRI}})$ and $\cos(\omega_0 m_\mathrm{s} T_{\mathrm{SRI}})$ up to second order under the small-rotation over the \ac{CPI}, i.e., $\omega_0 m_\mathrm{s} T_{\mathrm{SRI}} \ll 1$, \eqref{eq:R} can be rewritten as
\begin{align}\label{eq:R_espansione}
R_l&[{m_\mathrm{s}}]\approx R_\mathrm{c}[{m_\mathrm{s}}]
+ \bigl[x_l'\cos(\Psi_0)- y_l'\sin(\Psi_0)\bigr]\cos\varphi \\
& + z_l'\sin\varphi+ \underbrace{\bigl[-x_l'\sin(\Psi_0) - y_l'\cos(\Psi_0)\bigr]\,\omega_0 {m_\mathrm{s}}T_{\mathrm{SRI}} \cos\varphi}_{\Delta R_{\text{walk}}} \nonumber\\
& {}+ \underbrace{\bigl[-x_l'\cos(\Psi_0) + y_l'\sin(\Psi_0)\bigr]\tfrac{1}{2}\,\omega_0^2 ({m_\mathrm{s}}T_{\mathrm{SRI}})^2 \cos\varphi}_{\Delta R_{\text{curv}}}\nonumber
\end{align}
where $\Psi_0 = \Theta_0 - \theta_0$ is the initial aspect angle. From \eqref{eq:R_espansione}, it can be noticed that target rotation generated by the motion leads scatterer $l$ to migrate from its initial range bin due to two range-drift terms, denoted by $\Delta R_{\text{walk}}$ and $\Delta R_{\text{curv}}$ and referred to as range walk and range curvature, respectively.\footnote{It is worth noting that, when the combined effect of $\Delta R_{\text{walk}}$ and $\Delta R_{\text{curv}}$ produces a range migration that exceeds the range-resolution cell, the energy of $l$ spreads over multiple range bins and the image becomes defocused. In such cases, range-cell migration correction (RCMC) is required~\cite{chen2014inverse}.} 

From \eqref{eq:R_espansione}, the Doppler shift term can then be obtained by computing the derivative, as follows 
%
%
\begin{align}\label{eq:Doppler_freq}
f_D[{m_\mathrm{s}}] & = \frac{2}{\lambda}\,\frac{d R_l(t)}{d t}\bigg|_{t=m_\mathrm{s}T_\mathrm{SRI}}
 = \frac{2}{\lambda}\Biggl[\frac{d R_\mathrm{c}(t)}{d t}\bigg|_{t=m_\mathrm{s}T_\mathrm{SRI}}\Biggr. \\ 
&{} \hspace{1em} - \smash{\underbrace{\big[x_l'\sin(\Psi_0) + y_l'\cos(\Psi_0)\big]}_{u}}\,\omega_0 \cos\varphi \nonumber\\
& {} \Biggl.\hspace{1em} - \big[x_l'\cos(\Psi_0) - y_l'\sin(\Psi_0)\big]\,\omega_0^{2} {m_\mathrm{s}}T_{\mathrm{SRI}}\cos\varphi \Biggr]\nonumber
\end{align}

where the term denoted by $u$ represents the cross-range coordinate associated with scatterer $l$, while the quadratic Doppler term is responsible for cross-range distortions. However, the latter can be neglected when its amplitude is smaller than the Doppler frequency resolution $\Delta f_\mathrm{D} = 1/(M_\mathrm{s}T_{\mathrm{SRI}})$ over the entire \ac{CPI} \cite{CataldoPhD2016}. The corresponding cross-range resolution is
\begin{equation}
\begin{split}
\Delta u \;=\; \frac{\lambda}{2\,\omega_0 \cos \varphi}\,\Delta f_D.
\end{split}
\label{eq:Deltau}
\end{equation}
%


\section{Image Formation}\label{sec:image_formation}
For \ac{IVA} range–Doppler image formation, \ac{TMC} is required to estimate and compensate for the target’s time-varying translational motion parameters. These parameters are represented by the range shift terms $v_0 m_\mathrm{s} T_\mathrm{SRI}$ and $\tfrac{1}{2} a_0 (m_\mathrm{s} T_\mathrm{SRI})^2$ in $R_\mathrm{c}[m_\mathrm{s}]$, and its derivative given in \eqref{eq:Doppler_freq}. After compensation, the target range becomes time-invariant, and the Doppler frequencies remain approximately constant over the entire \ac{CPI}. Specifically, \ac{TMC} comprises both range alignment and phase adjustment. 

\subsection{Range alignment via cross-correlation}
The cross-correlation approach for range alignment detailed in \cite[Chapter 4]{chen2014inverse} is adopted. Starting from the frequency-domain sensing matrix $\mathbf{G}_\mathrm{s}$ in \eqref{eq:G_matrix}, a column-wise \ac{FFT} is first performed across subcarriers to obtain a matrix $\mathbf{F}_\mathrm{s}$, whose columns correspond to time-reversed range profiles. As a second step, an element-wise modulus operation is applied to $\mathbf{F}_\mathrm{s}$, and an \ac{IFFT} is then performed for each matrix column, i.e., \ac{OFDM} symbol. The reference profile is then selected as the column corresponding to sensing symbol index $(M_\mathrm{s}/2)-1$, and its complex conjugate is multiplied element-wise with each column of $\mathbf{F}_\mathrm{s}$, producing a correlation matrix $\mathbf{P}_\mathrm{s}$. An \ac{IFFT} across the frequency dimension (i.e., across rows) of $\mathbf{P}_\mathrm{s}$ yields the cross-correlation functions, whose peak locations indicate the range shifts (in range-cell units) relative to the reference profile. As a last operation, each range profile of index $m_\mathrm{s}$ can be shifted back by the related estimated shift amount ${q}_{m_\mathrm{s}} \in [-K_\mathrm{s}/2,(K_\mathrm{s}/2)-1]$. 
%
%

Each ${q}_{m_\mathrm{s}}$ is then mapped to a phase term as $\Phi_{m_\mathrm{s}} = \frac{2\pi}{K_\mathrm{s}}{q}_{m_\mathrm{s}}$. A phase unwrapping operation is subsequently applied to remove discontinuities larger than $\pi$, obtained by adding appropriate multiples of $\pm 2\pi$ so that adjacent phase differences satisfy $|\bar{\Phi}_{m_\mathrm{s}} - \bar{\Phi}_{m_\mathrm{s}-1}| < \pi$. The resulting continuous phase term $\bar{\Phi}_{m_\mathrm{s}}$ is finally mapped back to range-cell units as $\bar{q}_{m_\mathrm{s}} = \frac{K_\mathrm{s}}{2\pi}\bar{\Phi}_{m_\mathrm{s}}$.
Subsequently, a quadratic least-squares fit is performed to regularize the drift, as follows:
\begin{equation}
 (\tilde{a}_0,\tilde{a}_1,\tilde{a}_2)
= \arg\min_{a_0,a_1,a_2}
\sum_{j=1}^{M_s} \Big( \bar{q}_{j} - (a_0 + a_1 j + a_2 j^2) \Big)^2
\end{equation}
thereby obtaining the range drift $\tilde {q}_{m_\mathrm{s}} = (\tilde{a}_0 + \tilde{a}_1 {(m_\mathrm{s}+1)} + \tilde{a}_2 {(m_\mathrm{s}+1)}^2)$, with ${m_\mathrm{s}}=0,\ldots,M_\mathrm{s}-1$. 

%
%

The corresponding delay is then obtained as:
%
  ${\tau}_{m_\mathrm{s}} = \tilde{q}_{m_\mathrm{s}}\,\Delta \tau$, 
%
%
%
%
where $\Delta\tau = 1/({K_\mathrm{s} \Delta f})$ denotes the delay-domain sampling interval, corresponding to the system’s delay resolution.\footnote{The relationship between delay resolution and range resolution is given by $\Delta r = c \Delta \tau/2$.} 

Lastly, for each sensing sample $g_{k,m_\mathrm{s}}$ in \eqref{eq:G_matrix}, delay compensation is applied via a phase ramp across subcarriers ${k}$, thus obtaining 
  $\Gamma_{k,m_\mathrm{s}} \;=\; g_{k, m_\mathrm{s}}
  e^{\imath 2 \pi k \Delta f \tau_{m_\mathrm{s}}}$.
Given the matrix $\boldsymbol{\Gamma}$ of elements $\Gamma_{k,m_\mathrm{s}}$, $M_\mathrm{s}$ aligned range profiles $\mathbf{s}^{(m_\mathrm{s})}_\mathrm{R} \in \mathbb{C}^{K_p \times 1}$ are then reconstructed by performing a column-wise $K_\mathrm{p}$-point \ac{FFT} across subcarriers\footnote{When computing range profiles via \ac{FFT}, the result is a time-reversed and circularly shifted version of the \ac{IFFT}-based profile (see, e.g., \cite{Braun}). A one-bin circular rotation and a left–right flip are thus applied to the resulting profile to correctly position the zero-delay bin and obtain an increasing range axis.} 
\begin{equation}
  \mathbf{s}_{\mathrm{R}}^{({m_\mathrm{s}})} = \operatorname{FFT}_{K_\mathrm{p}}\!\left\{\boldsymbol{\Gamma}_{:,{m_\mathrm{s}}}\right\}
\end{equation}
where $K_\mathrm{p}> K_\mathrm{s}$ enables zero-padding in the frequency domain and is chosen as the smallest power of two greater than or equal to $2{K_\mathrm{s}}$.

\subsection{Minimum-variance phase-adjustment method}
\label{sec:phase-adjustment}
After range alignment, to remove phase drifts and obtain an approximately linear phase in the range cells occupied by the target, we apply a phase-adjustment procedure based on a minimum-variance criterion \cite[Chapter 4]{chen2014inverse}. Specifically, the minimum-variance method selects a small set of range cells whose envelope range profiles exhibit the lowest variance. Using these reference cells, we estimate a symbol-to-symbol phase-correction function, computed across symbols and averaged over the selected range cells, and then apply this correction to all range cells. 
In particular, the following \ac{CPE} function needs to be estimated 
\begin{equation}
\phi_{\mathrm{CPE}}(m_\mathrm{s}) \;=\; -\frac{4\pi}{\lambda}\,[ R_0 + v_0 m_\mathrm{s}T_{\mathrm{SRI}} + \tfrac{1}{2} a_0 (m_\mathrm{s}T_{\mathrm{SRI}})^2].
\end{equation}
%

Let $\mathcal{R}$ denote the set of $n_{\text{ref}}$ range cells (indexes) selected by the
minimum–variance criterion. A consistent estimate of the \ac{CPE} is
\begin{equation}
\widehat{\phi}_{\mathrm{CPE}}(m_\mathrm{s})
\!=\!\frac{1}{n_\mathrm{ref}}\!
\sum_{r\in\mathcal{R}}
\mathrm{unwrap}\big(\phi_r(m_\mathrm{s})-\phi_{r_0}(m_\mathrm{s})\big)
+\phi_{r_0}(m_\mathrm{s})
\label{eq:cpe-est}
\end{equation}
where $\phi_r(m_\mathrm{s})=\arg\{s_{\mathrm{R}}^{(m_\mathrm{s})}[r]\}$ and $r_0\in\mathcal{R}$ is the first reference cell. The  $\mathrm{unwrap(\cdot)}$ operator removes the $2\pi$ phase ambiguities as already explained.
The phase correction is then applied element-wise to all range cells at time index $m_\mathrm{s}$, as
\begin{equation}
\tilde{s}_{\mathrm{R}}^{(m_\mathrm{s})}[n] \;=\; s_{\mathrm{R}}^{(m_\mathrm{s})}[n]\,
e^{-\imath\,\widehat{\phi}_{\mathrm{CPE}}(m_\mathrm{s})},
\,\,\, n=0,\dots,K_\mathrm{p}-1.
\label{eq:phase-corr}
\end{equation}
%

%
%

From \eqref{eq:phase-corr}, by fixing the range cell $n$ and moving over the slow-time, we obtain the row vector $\mathbf{\tilde{s}}_\mathrm{R}[n]$. Finally, the range-Doppler image is formed by applying an $M_\mathrm{p}$-point \ac{FFT} to $\mathbf{\tilde{s}}_{\mathrm{R}}[n]$ over the $M_\mathrm{s}$ slow-time samples as
\begin{equation}
  \boldsymbol{\mathcal{P}}[n,:] \;=\; \left\lvert \operatorname{FFT}_{M_\mathrm{p}}\!\left\{\,\mathbf{\tilde{s}}_{\mathrm{R}}[n]\,\right\}\right\rvert, \quad n=0, \dots, K_\mathrm{p}-1
  \label{eq:I}
\end{equation}
where $M_\mathrm{p}>M_\mathrm{s}$ is here set to the smallest power of two greater than or equal to $10M_\mathrm{s}$, and $\boldsymbol{\mathcal{P}}[n,:]$ denotes the $n$th row of the range-Doppler matrix $\boldsymbol{\mathcal{P}} \in \mathbb{R}^{K_\mathrm{p} \times M_\mathrm{p}}$, whose columns are indexed by $q$.

\subsection{Imaging performance metrics}
The \ac{IC} quantifies image focusing and is defined as the normalized standard deviation of the image intensity \cite{chen2014inverse}. Given the IVA image data $\boldsymbol{\mathcal{P}}$ in \eqref{eq:I}, we compute IC over a region of interest $\widehat{\boldsymbol{\mathcal{P}}}\!\in\!\mathbb{R}^{\hat{K}_{\mathrm{p}}\times \hat{M}_{\mathrm{p}}}$ cropped from $\boldsymbol{\mathcal{P}}$ and centered at the ground-truth target-centroid. Precisely, letting $\mu$ denote the mean over all elements of $\widehat{\boldsymbol{\mathcal{P}}}$, i.e., $\mu =\frac{1}{\hat{M}_\mathrm{p} \hat{K}_\mathrm{p}}\mathbf{1}_{\hat{K}_\mathrm{p}}^\mathsf{T} \widehat{\boldsymbol{\mathcal{P}}} \mathbf{1}_{\hat{M}_\mathrm{p}}$,
%
%
the constrast is expressed as $\mathrm{IC}
=\frac{1}{\mu \sqrt{\hat{M}_\mathrm{p}\hat{K}_\mathrm{p}}}\|\widehat{\boldsymbol{\mathcal{P}}}-\mu \mathbf{1}_{\hat{K}_\mathrm{p}\times \hat{M}_\mathrm{p}} \|_F$.
%
%
%
%

In addition to \ac{IC}, system performance is also evaluated in terms of target-centroid estimation along the range dimension.
The image is first normalized to unit peak, $\tilde{\mathcal{P}}[n,q]=\mathcal{P}[n,q]/\max_{n,q} \mathcal{P}[n,q]$, then we define a threshold 
\begin{equation}
\delta \;=\; \epsilon\ \cdot \left(\mathcal{P}_{\max}-\mathcal{P}_{\min}\right)
\end{equation}
where $\mathcal{P}_{\max}=\max_{n,q}\tilde{\mathcal{P}}[n,q]$, $\mathcal{P}_{\min}=\min_{n,q}\tilde{\mathcal{P}}[n,q]$ and $\epsilon\in(0,1)$ sets the detection threshold as a fraction of the amplitude span. 
Then, 
the thresholded image is
\begin{equation}
\tilde{\mathcal{P}}_{\delta}[n,q] \;=\;
\begin{cases}
\tilde{\mathcal{P}}[n,q], & \text{if} \,\,\,\, \tilde{\mathcal{P}}[n,q] \ge \delta,\\[2pt]
0,            & \text{otherwise.}
\end{cases}
\end{equation}
The target-centroid range is finally estimated from $\tilde{\mathcal{P}}_{\delta}[n,q]$ as the intensity-weighted centroid along the range axis.

\section{Numerical Results}\label{sec:results}
This section analyzes the performance of the proposed monostatic \ac{ISAC}--\ac{IVA} setup.  For both communication and sensing, the \ac{BS} transmits 5G NR signals in the upper mid-band using \ac{QPSK}-modulated symbols, with $f_\mathrm{c}=6.7\,$GHz, $\Delta f=30\,$kHz and $K=13200$ active subcarriers, corresponding to a total bandwidth of around $400\,$MHz. The OFDM symbol duration and the frame duration are set to $T_\mathrm{s}= 35.7\,\mu$s, and $T_\mathrm{f}=10\,$ms, respectively, considering $M = 280$ \ac{OFDM} symbols per frame. The fraction of active subcarriers for sensing $\rho_\mathrm{f}$ is varied from $0.2$ to $1$, yielding a range resolution that spans from $\Delta r \approx 1.89\,$m to $\Delta r \approx 0.38\,$m, respectively. For the target, we consider two heading direction angles $\Theta_0\in\{270^{\circ},\,300^{\circ}\}$, measured anticlockwise from the positive $x$-axis. In this setup, the number of sensing symbols is fixed to $M_\mathrm{s}=220$ for heading $\Theta_0=270^\circ$ and $M_\mathrm{s}=200$ for $\Theta_0=300^\circ$. Using different $M_\mathrm{s}$ values is motivated by the need to keep $\omega_0\cos\varphi$ approximately constant over the \ac{CPI}. In both cases, the sensing repetition interval is $T_{\mathrm{SRI}}=1\,$ms, thus reserving only a fraction $\rho_\mathrm{t} \simeq 0.036$ of the frame-time resources for sensing. Moreover, $P_\mathrm{T} = 30\,$dBm, while $N_0=k_B T_0 F$, with  $k_B=1.38\cdot10^{-23}\,$J/K  the Boltzmann constant, $T_0=290\,$K, and $F=5\,$dB the noise figure.

The \ac{BS} is placed at $(x_\mathrm{r}, y_\mathrm{r}, z_\mathrm{r}) = (0,0,25)\,$m in the global system with two separate \acp{ULA} with $N_{\mathrm{T}} = N_{\mathrm{R}} = 10$ elements and azimuthal aperture (i.e., beamwidth) $\Delta\Theta_{\mathrm{T}} = \Delta\Theta_{\mathrm{R}} = 30^{\circ}$. Following the 3GPP Release~19 specifications \cite{TR38}, the extended vehicular target is modeled with $L=5$ point scatterers located at the front, rear, roof, left, and right sides, each having a unit, constant \ac{RCS}, at coordinates $(2.5, 0, 0)$, $(-2.5, 0, 0)$, $(0, 0, 0)$, $(0, 1, 0)$, and $(0, -1, 0)\,$m in the local (target-fixed) coordinate system. The target center, coinciding with the roof scatterer, is represented in global coordinates with $(x_\mathrm{t_c}, y_\mathrm{t_c}, z_\mathrm{t_c})$, with a fixed height $z_\mathrm{t_c}=1.6\,$m. Along the target trajectory, for the two considered heading angles $\Theta_0$, the midpoint-trajectory position is set at $(x_\mathrm{t_c}, y_\mathrm{t_c})=(60, 0)\,$m. When $\Theta_0 = 270^\circ$, the coordinate $x_\mathrm{t_c}$ remains constant throughout the trajectory.

\begin{figure}[t]
    \centering
    \includegraphics[width=0.32\textwidth]{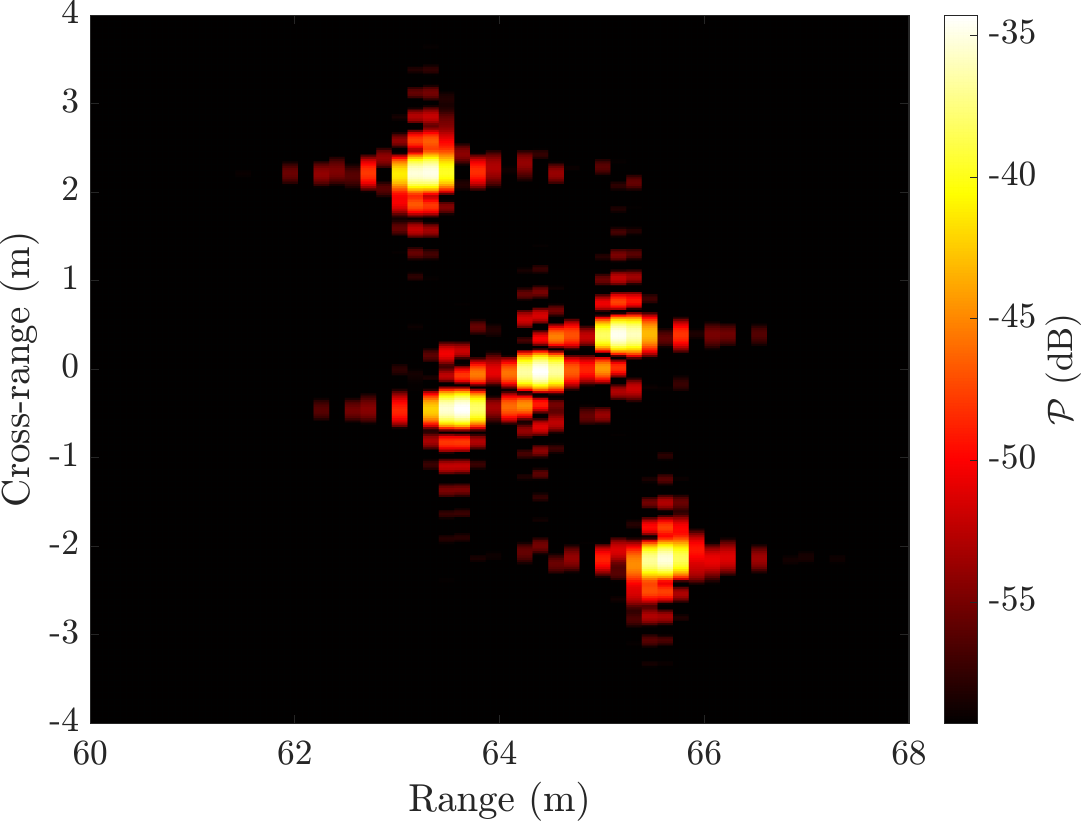}
    \caption{IVA image (\ac{IPP} plane) of the vehicle for $v = 30\,$m/s and $\rho_\mathrm{f} = 1$.}
    \label{fig:IVA_image}
\end{figure}

We consider three target speeds, $v \in \{10,20,30\}\,$m/s. The resulting cross-range resolution $\Delta u$, computed according to \eqref{eq:Deltau} is, respectively, $\Delta u \approx \{0.81,\,0.41,\,0.27\}\,$m for heading $\Theta_0 = 300^{\circ}$, and $\Delta u \approx \{0.66,\,0.33,\,0.22\}\,$m for $\Theta_0 = 270^{\circ}$. 
The number of reference range cells for phase-adjustment, selected via the minimum-variance criterion of Section~\ref{sec:phase-adjustment}, is set to $n_{\mathrm{ref}}=3$ for $\Theta_0=270^{\circ}$ and $n_{\mathrm{ref}}=5$ for $\Theta_0=300^{\circ}$, yielding the vehicle image shown in Fig. \ref{fig:IVA_image}.\footnote{Based on the target geometry, trajectory, and sensing parameters $M_\mathrm{s}$ and $T_{\mathrm{SRI}}$, the quadratic Doppler term, and the resulting range walk $\Delta R_{\mathrm{walk}}$ and range curvature $\Delta R_{\mathrm{curv}}$, are negligible and do not affect the imaging performance.}

Two analyses are conducted: 1) image-quality evaluation based on the \ac{IC} metric, and 2) assessment of the target-centroid estimation accuracy in the range domain. A total of $N_\mathrm{MC}=1000$ Monte Carlo iterations are performed along the target trajectory. For contrast analysis, $\hat{K}_\mathrm{p}$ and $\hat{M}_\mathrm{p}$ are chosen to match the dimensions of the image in Fig. \ref{fig:IVA_image}, and the mean image contrast $\mathrm{IC}_\text{mean}$ is obtained by averaging over $N_\mathrm{MC}$ iterations with $\Theta_0 = 300^\circ$. In the centroid analysis, accuracy is expressed through the \ac{RMSE} of the estimated centroid, denoted as $\mathrm{RMSE}_\mathrm{c}$, for both $\Theta_0$ values and a threshold of $\epsilon=0.3$.

\subsection{Image contrast}

\begin{figure*}[!t] 
\centering
\begin{minipage}[t]{0.455\textwidth}
    \centering
    \includegraphics[width=0.82\linewidth]{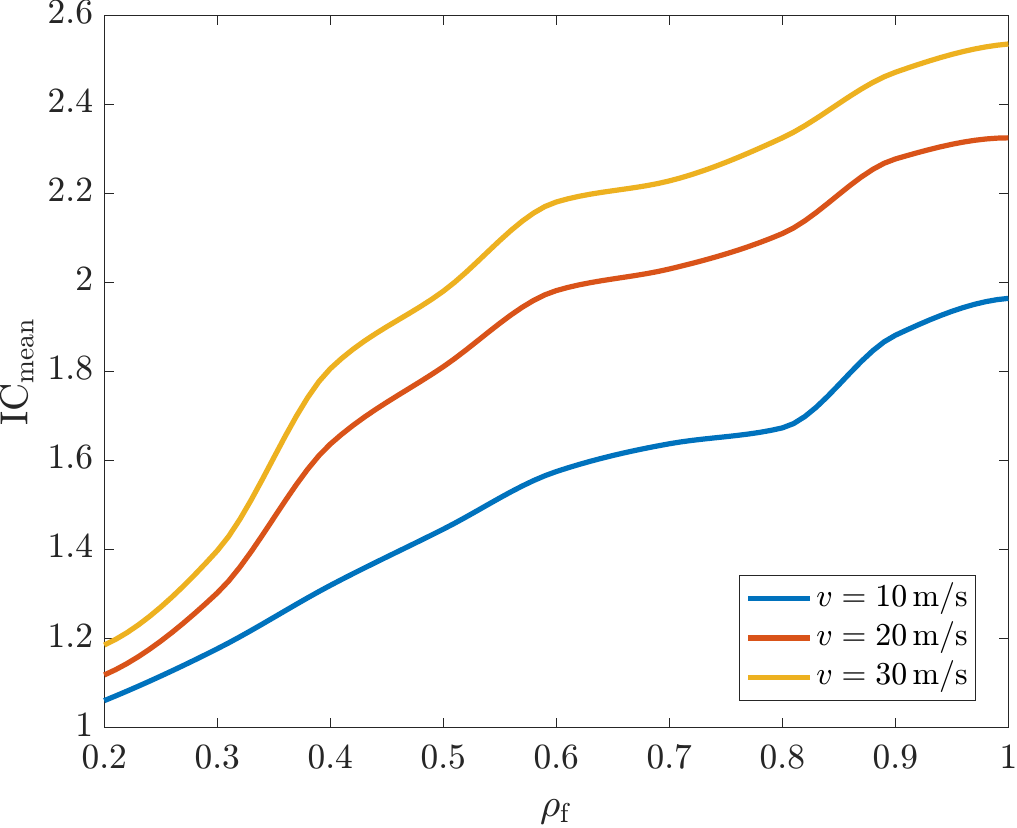}
    \caption{$\mathrm{IC}_{\text{mean}}$ of the \ac{IVA} image for $\Theta_\mathrm{0} = 300^\circ$ as a function of target speed and sensing subcarrier fraction $\rho_\mathrm{f}$.}
    \label{fig:IC_mean}
\end{minipage}%
\hfill
\begin{minipage}[t]{0.455\textwidth}
    \centering
    \includegraphics[width=0.81\linewidth]{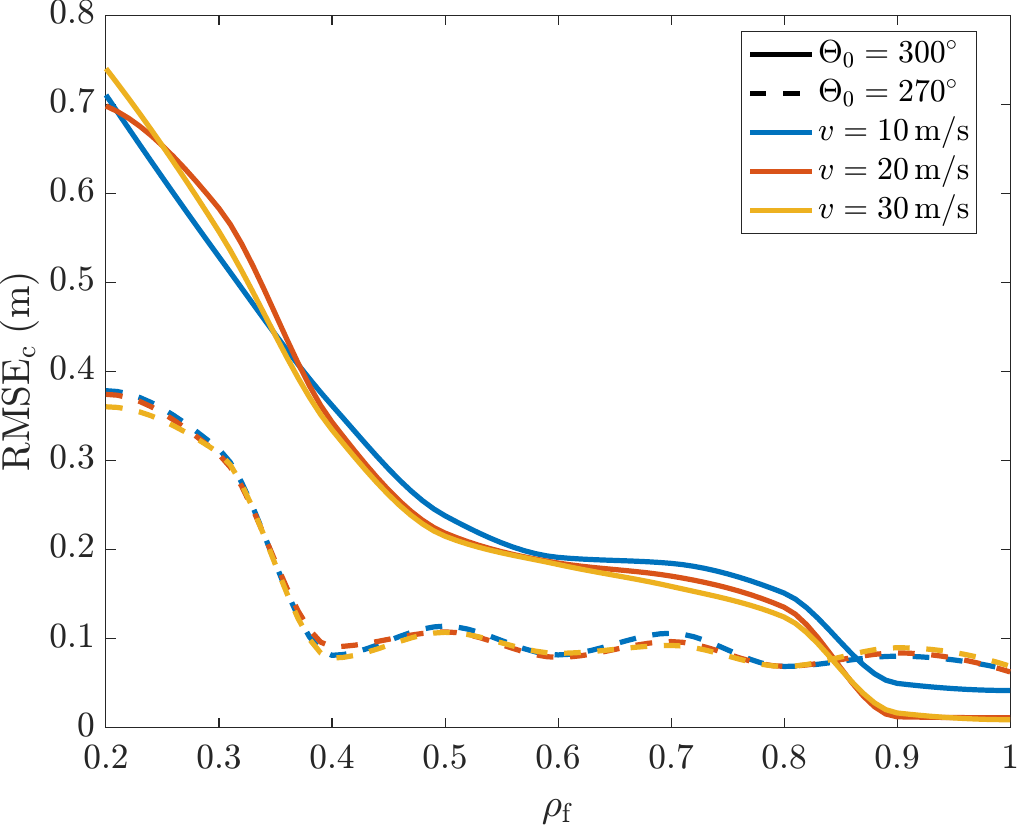}
    \caption{Target–centroid range estimation error as a function of target speed and sensing subcarrier fraction $\rho_{\mathrm f}$.}
    \label{fig:RMSE}
\end{minipage}
\end{figure*}

Fig.~\ref{fig:IC_mean} shows $\mathrm{IC}_\text{mean}$ versus the fraction of subcarriers allocated to sensing for three target speeds and $\Theta_0 = 300^\circ$.
While the cross-range resolution $\Delta u$ is fixed for a given speed, the range resolution improves with increasing $\rho_f$, which explains the overall upward trend of the $\mathrm{IC}_\text{mean}$ curves.
For $\rho_\mathrm{f} \leq 0.3$, the images remain blurred, and the target shape is barely recognizable. When $\rho_\mathrm{f}$ equals $0.7$, $0.6$, and $0.5$ at target speeds of $10$, $20$, and $30$ m/s, respectively, the target shape starts to become clearly discernible, with the roof, left, and right point scatterers appearing as distinct features. Beyond these operating points, further increasing $\rho_\mathrm{f}$ provides additional image contrast; however, a parsimonious choice of $\rho_\mathrm{f}$ preserves bandwidth for communication while still ensuring satisfactory target representation and recognition. 
Finally, higher target speeds yield larger $\mathrm{IC}_\text{mean}$ values, consistent with the fact that higher translational velocity induces a larger effective angular velocity and thus finer cross-range resolution.


\subsection{\ac{RMSE} of target-centroid range}
Fig.~\ref{fig:RMSE} shows the target-centroid range $\mathrm{RMSE}_\mathrm{c}$ varying the fraction of subcarriers for sensing. 
For $\Theta_0 = 270^{\circ}$, $\mathrm{RMSE}_\mathrm{c}$ decreases rapidly as $\rho_\mathrm{f}$ increases, while for $\Theta_0 = 300^{\circ}$ the reduction is more gradual and starts from larger  errors for small $\rho_\mathrm{f}$. 
This difference arises from the target geometry: at $\Theta_0 = 270^{\circ}$, the front and rear scatterers are nearly iso-range with the roof point, thus stabilizing the centroid even under coarse range resolution. Contrariwise, this geometric alignment is absent for $\Theta_0 = 300^{\circ}$.
As $\rho_\mathrm{f}$ increases and range resolution improves, both trajectories achieve very low errors, with $\mathrm{RMSE}_\mathrm{c} < 0.1\,$m for $\rho_\mathrm{f} > 0.8$. Notably, for $\Theta_0 = 270^\circ$, comparable estimation accuracy is already achieved for moderate values of  $\rho_\mathrm{f}\approx0.4$, showing that reliable centroid localization can be maintained while preserving bandwidth for communication. 
At very high $\rho_\mathrm{f}$, the $\Theta_0 = 300^{\circ}$ case slightly outperforms due to its centroid lying closer to the center of a sampled range bin at the attained resolution. In both cases, target speed (namely cross-range resolution) has only a marginal effect compared with $\rho_\mathrm{f}$ variations.


\section{Conclusion}\label{sec:conclusion}
In this work, an \ac{IVA} technique for a monostatic \ac{ISAC} system using a \ac{MIMO}-\ac{OFDM} waveform was developed, and its performance was evaluated for a moving extended vehicular target. Range alignment and phase compensation were performed via known cross-correlation and the minimum-variance method. Simulation results show that increasing the sensing subcarrier fraction $\rho_\mathrm{f}$ improves image focusing and reduces the target-centroid range estimation error, with the target shape becoming discernible and $\mathrm{RMSE}_\mathrm{c}\!<30\,$cm for $\rho_\mathrm{f}\geq 0.5$ across both considered trajectories. These findings indicate that good target localization and high-quality imaging can be achieved even with a moderate sensing bandwidth, while using only $3.6\%$ of the frame-time for sensing, enabling an effective trade-off between sensing performance and communication resource utilization.

%

\balance
\bibliographystyle{Monostatic_ISAC_ISAR/bibliography/IEEEtran}
\bibliography{Monostatic_ISAC_ISAR/bibliography/IEEEabrv,Monostatic_ISAC_ISAR/bibliography/Bibliography}


\end{document}

%% file: acronyms.tex
\begin{acronym}
\acro{AWGN}{additive white Gaussian noise}
\acro{AoA}{angle of arrival}
\acro{AoD}{angle of departure}
\acro{BF}{beamformer}
\acro{BS}{base station}
\acro{CP}{cyclic prefix}
\acro{CPE}{common phase error}
\acro{CPI}{coherent processing interval}
\acro{DoA}{direction of arrival}
\acro{DoD}{direction of departure}
\acro{EIRP}{effective isotropic radiated power}
\acro{ELP}{equivalent low-pass}
\acro{FFT}{fast Fourier transform}
\acro{GOSPA}{Generalized Optimal Sub-Pattern Assignment}
\acro{ISAR}{inverse synthetic aperture radar}
\acro{ISAC}{integrated sensing and communication}
\acro{ISI}{intersymbol interference}
\acro{ICI}{intercarrier interference}
\acro{IFFT}{inverse fast Fourier transform}
\acro{i.i.d.}{independent, identically distributed}
\acro{IVA}{inverse virtual aperture}
\acro{IPP}{image projection plane}
\acro{IC}{image contrast}
\acro{JSC}{joint sensing and communication}
\acro{LoS}{line-of-sight}
\acro{LTE}{long term evolution}
\acro{MCL}{maximum coupling loss}
\acro{MPL}{maximum path loss}
\acro{MIL}{maximum isotropic loss}
\acro{MC}{Monte Carlo}
\acro{MIMO}{multiple-input multiple-output}
\acro{MUSIC}{MUltiple SIgnal Classification}
\acro{MDL}{minimum description length}
\acro{NR}{new radio}
\acro{OFDM}{orthogonal frequency division multiplexing}
\acro{OSPA}{optimal sub-pattern assignment}
\acro{PSD}{power spectral density}
\acro{QPSK}{quadrature phase shift keying}
\acro{RCS}{radar cross-section}
\acro{RF}{radio frequency}
\acro{RMSE}{root mean squared error}
\acro{r.v.}{random variable}
\acro{SRI}{sensing repetition interval}
\acro{SSIR}{signal-to-self interference ratio}
\acro{SI}{self interference}
\acro{SNR}{signal-to-noise ratio}
\acro{SU}{single-user}
\acro{SCM}{sample covariance matrix}
\acro{SAR}{synthetic aperture radar}
\acro{TMC}{translational motion compensation}
\acro{UE}{user equipment}
\acro{ULA}{uniform linear array}
\acro{VA}{virtual aperture}
\end{acronym}